\documentclass[11pt,twoside]{article}
\usepackage[german,english]{babel}
\usepackage{a4,makeidx,fancyheadings}
\usepackage{pslatex}
\usepackage{cite}

\input epsf
\makeatletter
\@addtoreset{equation}{section}
\makeatother
\renewcommand\thesection{\Roman{section}}
\renewcommand\theequation{\ifnum \value{section}>0
 \thesection.\arabic{equation}%
\else
\arabic{equation}%
\fi}

\def\Re{\mbox{Re}}
\def\msbar{{\ensuremath{\overline{{\rm MS}}}}}

\newcommand{\xb}{\bar x}
\newcommand{\order}{{\cal O}}
\newcommand{\as}{\alpha_s}
\newcommand{\eq}[1]{eq.~(\ref{#1})}
\newcommand{\Li}[1]{\mbox{Li}_2\left(#1\right)}

\newcommand{\tr}{\mbox{\sf Tr}}
\newcommand{\st}[1]{\mbox{\scriptsize #1}}
\newcommand{\nn}{\nonumber}

\newcommand{\e}{\epsilon}
\newcommand{\T}{\Theta}

\newcommand{\ksl}{k\!\!\!/}

\newcommand{\rg}{r_\Gamma}

\newcommand{\smin}{s_{\st{min}}}

\def\L{\left(}\def\R{\right)}

\def\LB{\left[}\def\RB{\right]}

\def\qb{{\overline{q}}}
\def\Qb{{\overline{Q}}}

\def\Li#1{\,{{\rm Li}_{#1}}}
\def\xg{{x_g}}
\def\h6born{h_6^{\st{LO}} } 
 
\def\deltah6{\delta h_6}
\def\dRcoll{dR_{\st{coll.}}}
\def\dRsoft{dR_{\st{soft}}} 
\def\epem{{e^+e^-}} 
\def\Qb{{\ensuremath{\bar{Q}}}} 
\def\qb{{\bar q}}
\def\T{{\cal T}}
\def\cC{{\cal C}}

\def\pss{phase space slicing }

\begin{document}
\thispagestyle{empty}
hep-ph/0008291 \hfill{\parbox[t]{3.8cm}{DESY 00-121\\
    Saclay/SPhT--T00/120}}
\begin{center}
{\bf
Next-to-leading order QCD corrections to 
parity-violating 3-jet observables for massive quarks in
$\bf e^+e^-$ annihilation}  
\end{center}

\centerline{
W. Bernreuther$^{a,}$\footnote{{\tt breuther@physik.rwth-aachen.de},
research supported by BMBF, contract 05 HT9PAA 1}, 
A. Brandenburg$^{b,}$\footnote{{\tt arnd.brandenburg@desy.de},
research supported by
Heisenberg Fellowship of D.F.G.} 
and P. Uwer$^{c,}$\footnote{{\tt uwer@spht.saclay.cea.fr}}}

\begin{center}
$^a$Institut f\"ur Theoretische Physik,
RWTH Aachen, D-52056 Aachen, Germany.\\
$^b$DESY Theory Group, D-22603 Hamburg, Germany.\\
$^c$Service de Physique Th{\'e}orique,
Centre d'Etudes de Saclay, \mbox{F-91191 Gif-sur-Yvette cedex}, France.
\end{center}

\centerline{\bf Abstract:}
In this work we provide all the missing ingredients for calculating
parity-violating 3-jet observables in $e^+e^-$ collisions 
at next-to-leading order (NLO)
including the full quark mass dependence. In particular we give explicit
results for the one-loop corrections and for the singular 
contributions from real emission processes. 
Our formulae allow  for the computation of the
NLO 3-jet contributions to the forward-backward asymmetry
for heavy flavours, which is -- and will be -- an important observable for
electroweak physics at present and future $e^+e^-$ colliders.
%

%
%
\setcounter{page}{0}
\newpage
\begin{section}{Introduction}
The measurement of the forward-backward asymmetry $A_{\st{FB}}^b$ for 
$b$-quark production in
$e^+ e^-$ collisions at the Z peak provides one of the most precise
determinations of the weak mixing angle ${\rm sin}^2\theta_{\rm eff}$. It
can be determined with an error at the per mille level if the 
computations of $A_{\st{FB}}^b$
are at least as accurate as the present experimental 
precision of about two percent \cite{LEP99}.
This implies, in particular, that perturbative and 
non-perturbative QCD corrections to the lowest order formula
for $A_{\st{FB}}^b$  must be controlled to this level of accuracy.

As far as perturbative QCD corrections to this observable are concerned, 
the present knowledge is as follows.
The order $\alpha_s$ contributions had been computed first 
for massless and then for massive quarks in ref. \cite{JeLaZe81} 
and ref. \cite{DjKuZe90,ArBaLe92}, respectively. These calculations, which used 
the quark direction for defining the asymmetry 
were later modified
\cite{DjLaZe95,DjZe97,La96} to encompass the asymmetry with respect to the 
thrust axis, which is more relevant 
experimentally. The coefficient of the order $\alpha_s^2$ 
correction was computed first in ref. \cite{AlLa93}, for massless
quarks using the quark axis definition, by numerical
phase space integration. This result was recently corrected 
by a completely analytical \cite{NeRa98} and by a numerical
calculation \cite{CaSe99b}. 
In the latter reference the second-order corrections
were also determined for the thrust axis definition.

For future applications to measurements at proposed linear $e^+e^-$ colliders 
the attained theoretical precision so far will not suffice. 
The forward-backward asymmetry will be
a key observable for determining the
electroweak couplings of the top quark above threshold -- 
a fundamental task at a linear collider as far as top quark
physics is concerned. 
Needless to say all radiative corrections
must be determined for non-vanishing top quark mass.
Moreover, there may be the option to run 
such a collider at the Z peak, aiming to increase the precision attained at
LEP1/SLC significantly. For instance one may be able to
 measure $A_{\st{FB}}$ for $b$ and $c$ quarks to an
accuracy of order 0.1 $\%$. (For a recent discussion, see ref. \cite{HaMo99}.)
\par
Here we are concerned with the second-order QCD corrections to  $A_{\st{FB}}$
for massive quarks $Q$. It is convenient 
to obtain this asymmetry by computing the individual
contributions of the parton jets \cite{AlLa93}. 
This break-up may also be useful from
an experimental point of view, as the various contributions 
can be separately measured \cite{BuOs97,SLD00,Delphi00}.
To order $\alpha_s^2$ the asymmetry is determined by
the 2-, 3-, and 4-jet final states involving $Q$. 
The 4-jet contribution to $A_{\st{FB}}$ involves the known 
parity-violating
piece of the Born matrix elements for $e^+e^- \to 4$ partons 
involving at least one ${\bar Q}Q$ pair (see, e.g. ref. \cite{BaMaMo94}).
The phase-space integration of these matrix elements
in order to get $(A_{\st{FB}})_{\st{4-jet}}$ for a given jet-resolution algorithm can
be done numerically in a rather
straightforward fashion. 
The computation of $(A_{\st{FB}})_{\st{3-jet}}$ is highly non-trivial.
Within the phase space slicing method \cite{GiGl92,GiGlKo93,KeLa99}, 
which we will use here,  
it amounts to determining first the parity-violating
piece in the fully differential cross section for 
the reaction $e^+e^- \to 3$ {\it resolved partons} to order $\alpha_s^2$. 
This will be done in this paper.
(For the definition of `resolved partons' see section 
\ref{sec:realcontributions}.)
In addition the contribution to $(A_{\st{FB}})_{\st{3-jet}}$ from 
$e^+e^- \to 4$ resolved partons is
needed. 
It can be obtained from the known leading
order matrix elements by a numerical integration in 4 space-time
dimensions which incorporates a jet defining algorithm. 
This will be discussed elesewhere.
The computation of $(A_{\st{FB}})_{\st{2-jet}}$ for heavy quarks 
is beyond the scope of this work and remains to be done in the future.
\par
Our paper is organized as follows. In section \ref{sec:kinematics} we 
review the kinematics of the reaction
and leading-order results, and
define the 3-jet forward-backward asymmetry. 
In section \ref{sec:virtual-corrections} we  discuss the calculation of the
virtual corrections. In particular, the ultraviolet divergences,
the infrared divergences, and the renormalization procedure is discussed.
In section \ref{sec:realcontributions} the singular contributions from
the real corrections are calculated and the cancellation of the 
infrared divergences is checked. 
In section \ref{sec:conclusions} we summarize our results and end with
some conclusions.
\end{section}

\section{Kinematics and leading-order results}
\label{sec:kinematics}
\def\bp{{\bf p}}
\def\bk{{\bf k}}
The leading order differential cross section for
\begin{equation}
  \label{eq:QQbg}
  e^+(p_+)e^-(p_-)\to Q(k_Q) \bar{Q}(k_{\bar{Q}})g(k_g)
\end{equation}
 can be written in the following form{\footnote{
We neglect the lepton masses and do not consider transversely polarized
beams.}}: 
\begin{eqnarray}
\label{eq:dsigma}
  \frac{d\sigma}{d\phi d\!\cos\vartheta  dx d\xb }
  &=&
  \frac{3}{4}\frac{1}{(4\pi)^3}\sigma_{pt}\Bigg\{
  (1+\cos^2\vartheta)\, F_1 + (1-3\cos^2\vartheta)\,F_2 
  + \cos\vartheta\,F_3\nonumber\\
  &+& 
  \sin 2\vartheta \cos\phi\, F_4 
  + \sin^2\vartheta \cos 2\phi\,F_5
  + \sin\vartheta\cos\phi\,F_6
  \label{eq:diffcs}
  \Bigg\},
\end{eqnarray}
with 
\begin{equation}
  \label{eq:sigmapt}
  \sigma_{pt} = \sigma(e^+e^-\rightarrow \gamma^\ast 
  \rightarrow\mu^+\mu^-) = 
  \frac{4\pi\alpha^2}{3s}.
\end{equation}
In \eq{eq:dsigma}, $\vartheta$ denotes the angle between the 
direction ${\bf p}_-$ of the incoming electron $e^-$
and the direction ${\bf k}_Q$ 
of the heavy quark $Q$. The angle $\phi$ is the oriented 
angle between the plane defined by $e^-$ and $Q$
 and the plane defined by the quark anti-quark pair.
The functions $F_i$ depend only on the scaled c.m. energies 
$x=2k\cdot k_Q/s$ and $\xb=2k\cdot k_{\bar{Q}}/s$, and on 
the scaled mass square
$z=m^2/s$ of the heavy quark and anti-quark,
where $k=p_+ + p_-$ and $s=k^2$. Note that the leading order  
differential cross section for a final state with $n\ge 4$ 
partons can be written in an analogous 
way in terms of the two angles $\vartheta$,  $\phi$ and 
$3n-7$ variables that involve only the final state momenta.
In higher orders absorptive parts give rise to three additional 
functions $F_{7,8,9}$ (cf. ref. \cite{BrDiSh96}).
The four functions $F_{1,2,4,5}$ are parity even.
In particular, the 3-jet production rate is determined by $F_1$. 
In the case of massive quarks 
next-to-leading order results for $F_1$ were given
in references
\cite{BeBrUw97a,BrUw97,Ro96,Ro97,BiRoSa97b,BiRoSa99,NaOl97b, Ol97}.
The two functions $F_3$ and $F_6$ are induced by the interference
of a vector and an axial-vector current. In particular, 
using the quark-axis, the 3-jet 
forward-backward asymmetry is related to $F_3$ in leading order 
in the following way:
\begin{eqnarray}
\label{eq:defafb}
(A_{\st{FB}})_{\st{3-jet}} &=& {
\int_{0}^{1}d\cos\vartheta\int_{-\pi}^{\pi}d\phi
\int_{D(y_{\rm cut})}dx d\xb 
\frac{d\sigma}{d\phi d\!\cos\vartheta  dx d\xb } -
\int_{-1}^{0}d\cos\vartheta\int_{-\pi}^{\pi}d\phi
\int_{D(y_{\rm cut})}dx d\xb 
\frac{d\sigma}{d\phi d\!\cos\vartheta  dx d\xb}
\over
\int_{-1}^{1}d\cos\vartheta\int_{-\pi}^{\pi}d\phi
\int_{D(y_{\rm cut})}dx d\xb 
\frac{d\sigma}{d\phi d\!\cos\vartheta  dx d\xb} 
} \nonumber \\ 
&=& \frac{3}{8}
{\int_{D(y_{\rm cut})}dx d\xb F_3(x,\xb) \over 
\int_{D(y_{\rm cut})}dx d\xb F_1(x,\xb) },
\end{eqnarray}    
where $D(y_{\rm cut})$ defines, for a given jet finding algorithm 
and jet resolution parameter
$y_{\rm cut}$, a region in the $(x,\xb)$ plane.
Analogously, the function $F_6$ induces an azimuthal asymmetry
\begin{eqnarray}
\label{eq:F6asym}
A_{\phi} &=& {
\int_{0}^{\pi/2}d\phi\int_{-1}^{1}d\cos\vartheta
\int_{D(y_{\rm cut})}dx d\xb 
\frac{d\sigma}{d\phi d\!\cos\vartheta  dx d\xb } -
\int_{\pi/2}^{\pi}d\phi\int_{-1}^{1}d\cos\vartheta
\int_{D(y_{\rm cut})}dx d\xb 
\frac{d\sigma}{d\phi d\!\cos\vartheta  dx d\xb}
\over
\int_{0}^{\pi/2}d\phi\int_{-1}^{1}d\cos\vartheta
\int_{D(y_{\rm cut})}dx d\xb 
\frac{d\sigma}{d\phi d\!\cos\vartheta  dx d\xb } +
\int_{\pi/2}^{\pi}d\phi\int_{-1}^{1}d\cos\vartheta
\int_{D(y_{\rm cut})}dx d\xb 
\frac{d\sigma}{d\phi d\!\cos\vartheta  dx d\xb}
} \nonumber \\ 
&=& \frac{3}{8}
{\int_{D(y_{\rm cut})}dx d\xb F_6(x,\xb) \over 
\int_{D(y_{\rm cut})}dx d\xb F_1(x,\xb) }.
\end{eqnarray}
Note that final states with two partons do not contribute to
$A_{\phi}$.
The electroweak couplings appearing in $F_{3,6}$ can be factored
out as follows:
\begin{equation}
\label{eq:f36}
F_{3,6} = -g_a^Q(1-\lambda_+\lambda_-)\left[Q_Q {\rm Re} \chi (g_a^e
-f(\lambda_+,\lambda_-)g_v^e)
+g_v^Q |\chi|^2 (f(\lambda_+,\lambda_-)(g_v^{e2}+g_a^{e2})-2g_v^eg_a^e)
 \right]\tilde{F}_{3,6}.
\end{equation}
In \eq{eq:f36},
\begin{eqnarray}
g_v^f &=& T_3^f-2Q_f\sin^2\theta_W,\nonumber \\
g_a^f &=& T_3^f, \nonumber \\
\chi &=& {1\over 4\sin^2\theta_W\cos^2\theta_W}
{s\over s-m_Z^2+i m_Z\Gamma_Z}, \nonumber \\
f(\lambda_+,\lambda_-) &=& {\lambda_--\lambda_+\over 1-\lambda_-\lambda_+},
\end{eqnarray}
where $T_3^f$ is the third component of the weak isospin of the fermion $f$,
$\theta_W$ is the weak mixing angle, and  $\lambda_\mp$ denotes the 
longitudinal polarization of the electron (positron).
In next-to-leading order in $\alpha_s$, additional contributions to 
$F_{3,6}$ with electroweak couplings different from those 
in \eq{eq:f36} are induced which we
will not consider here. They are either proportional to ${\rm Im}\chi$
and thus suppressed formally in the electroweak coupling
or generated by the triangle fermion loop diagrams depicted 
in figure 1(l), 1(m). The contribution of the triangle fermion loop
is gauge independent and UV and IR finite. It was calculated some time ago 
in ref. \cite{HaKuYa91}.

The functions $\tilde{F}_{3,6}(x,\xb)$ may be expressed 
in terms of  functions $h_6$, $h_7$ which appear in the 
decomposition of the so-called hadronic tensor 
as performed for example in references \cite{KoSc85,KoSc89,KoScKrLa86}.
\begin{eqnarray}
\label{eq:h6f36}
\tilde{F}_3 &=& {1\over 2}\left[\sqrt{x^2-4z}h_6(x,\xb)
+\sqrt{\xb^2-4z}\cos\vartheta_{Q\bar{Q}}h_7(x,\xb)\right],\nonumber \\
\tilde{F}_6 &=& -{1\over 2}\sqrt{\xb^2-4z}\sin\vartheta_{Q\bar{Q}}h_7(x,\xb),
\end{eqnarray}
where $\vartheta_{Q\bar{Q}}$ is the angle between $Q$ and $\bar{Q}$ in the
c.m. frame and we have
\begin{equation}
\cos\vartheta_{Q\bar{Q}}={2(1-x-\xb+2z)+x\xb
\over \sqrt{x^2-4z}\sqrt{\xb^2-4z}}.
\end{equation}
It would seem pointless to trade $F_{3,6}$ for $h_{6,7}$ if it were not
for the relation (which follows from CP invariance):
\begin{equation}
\label{eq:h7}
h_7(x,\xb)=-h_6(\xb,x).
\end{equation}  
The objective of this paper is to provide explicit
expressions for the virtual corrections
and for the singular parts of the real corrections to the function
$h_6$. These corrections make up the complete
contribution from  three {\it resolved partons}
at NLO, $h_6^{\st{3 res., NLO}}$, which is ultraviolet and
infrared finite.
We regulate both infrared und ultraviolet singularities 
by  continuation to $d=4-2\e$ space-time 
dimensions. 
In this context one may wonder whether the above kinematics relations, in
particular \eq{eq:h6f36} have to be modified for $d\not=4$.
This is however not necessary if we consistently keep the momenta
of electron and positron as well as the polarization vector
of the photon/Z-boson in $d=4$ dimensions throughout the 
calculation. Note that this prescription still corresponds to 
the usual dimensional regularisation. In addition this prescription
is compatible with the `t~Hooft-Veltman prescription for $\gamma_5$ as we
discuss later.
 
We finish this section by giving  
the Born result for $h_6$ in $d=4-2\e$ dimensions:
\begin{eqnarray}
  \label{eq:bornresult}
 \h6born(x,\xb) 
 = 
 16\*\pi\*\as\*(N^2-1) \* B \*
 \bigg(2\* x
 -(x\*\xb+\xb^2+2-4\*\xb)\*\e
 -4\*{\xg\over 1-x}\*z
 \bigg)
\end{eqnarray}
with 
\begin{equation}
  \label{eq:Bdef}
  B = {1\over (1-x)\*(1-\xb)}
  ,
\end{equation}
and the scaled gluon energy
\begin{equation}
  x_g = {2k\cdot k_g\over s}= 2-x-\xb.
\end{equation}
Note that the terms proportional to $\e$ in \eq{eq:bornresult} depend on the
prescription used to treat $\gamma_5$ in $d$ dimensions. To derive the
above equation we have used the prescription 
$ \gamma_\mu \gamma_5\rightarrow 
  {i\over3!}\varepsilon_{\mu\beta\gamma\delta}\,
  \gamma^\beta\gamma^\gamma\gamma^\delta
$ \cite{tHoVe72,La93}.
We will discuss this issue in more detail in the next section together
with the ultraviolet (UV) renormalization.

\section{Virtual corrections}
\label{sec:virtual-corrections}
\begin{figure}[htbp]
  \begin{center}
    \leavevmode 
    \centerline{\epsfxsize9cm\epsfbox{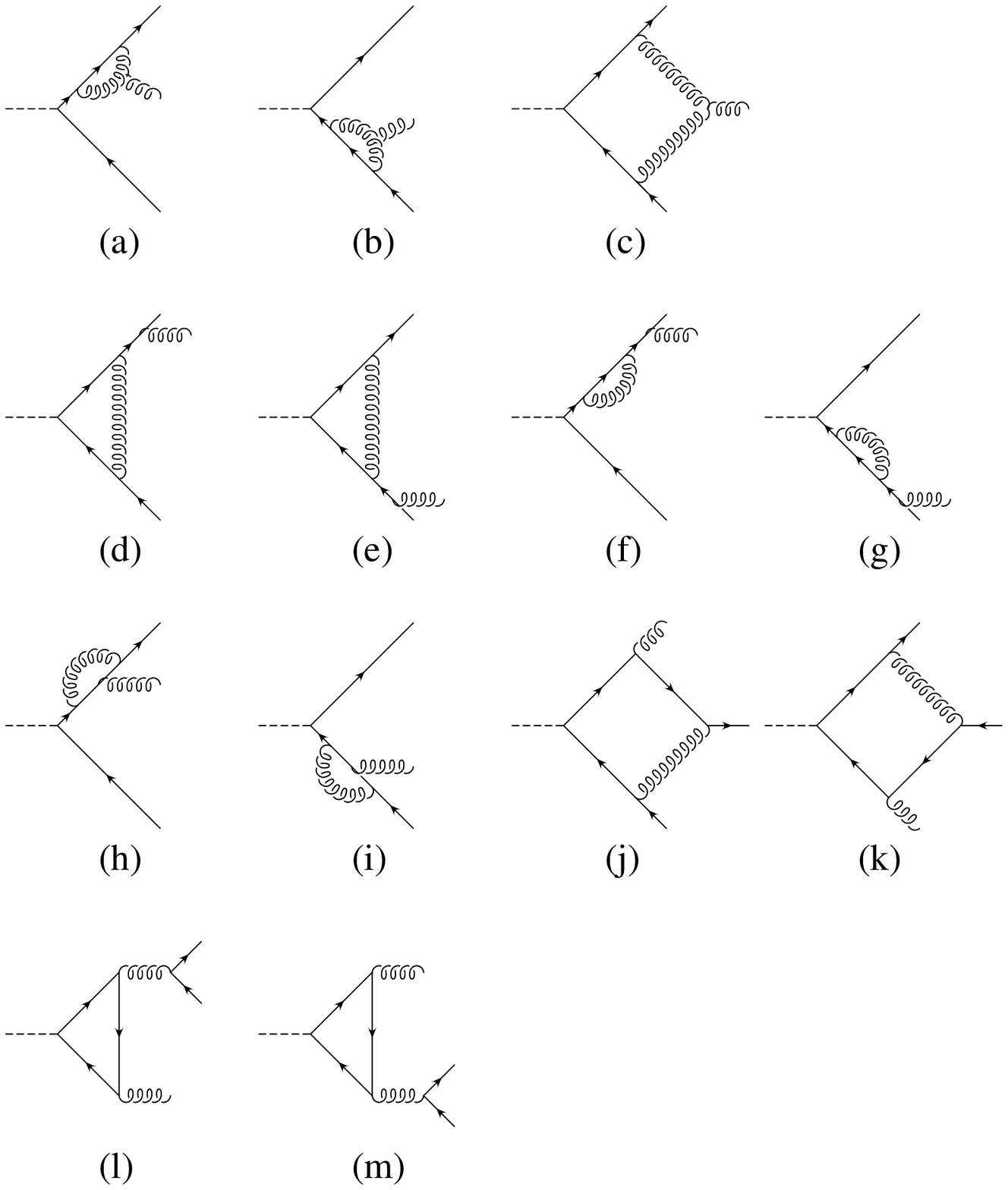}}
    \caption{One-loop diagrams to $Z\to Q\Qb g$.}
    \label{fig:loop-diagrams}
  \end{center}
\end{figure}
In this section we will give analytic expressions for the 
virtual corrections to $h_6$, isolate the ultraviolet
and infrared singularities, carry out the renormalization
procedure, and discuss the treatment of $\gamma_5$.

We work in renormalized perturbation theory, which  means that the
bare quantities (fields and couplings) are expressed
in terms of renormalized quantities. By this procedure one obtains
two contributions: one is the original Lagrangian but now in terms of
the renormalized quantities, the second contribution are the so-called 
counterterms:
\begin{equation}
  \label{eq:ren-pert}
  {\cal L}(\Psi_0,A_0, m_0,g_0) = 
  {\cal L}(\Psi_R,A_R, m_R,g_R) + {\cal L}_{ct}(\Psi_R,A_R, m_R,g_R).
\end{equation}
The first contribution yields the same Feynman rules as the bare
Lagrangian but with the bare quantities replaced by renormalized ones.
In the following we renormalize the quark field and the quark mass 
in the on-shell scheme,  but treat the gluon field and the strong coupling 
in the modified minimal 
subtraction (\msbar) scheme \cite{tHo73,BaBuDuMu78}, taking into account 
the Lehmann--Symanzik--Zimmermann residue in the S-matrix element.
Note that the conversion of the on-shell mass to the frequently used
\msbar\ mass can be performed at the
end of the calculation. The quark mass in the on-shell scheme
will be denoted in the following  simply  by $m$ for notational
convenience.

While in the \msbar\ scheme the renormalization constants contain
only UV singularities, in the on-shell scheme they contain also
infrared (IR) 
divergences\footnote{
Note that the cross section at hand is inclusive enough so that no
uncancelled collinear singularities survive. For simplicity we thus
call soft and collinear singularities just IR singularities.}. 
Although at the very end all the divergences cancel
it is worthwhile to distinguish between UV and IR singularities so that
one can check the UV renormalization and the IR finiteness independently.

To start with let us discuss the contribution of the one-loop diagrams
before renormalization, that is the contribution from  
${\cal L}(\Psi_R,A_R, m_R,g_R)$.
Although tedious the calculation of the one-loop diagrams shown
in figure \ref{fig:loop-diagrams} is in principle straightforward.
We do not discuss here the contribution from diagrams
 l) and m) as mentioned earlier. 
Without going into the details we just review some techniques used in 
the calculation of the remaining diagrams. 
To calculate the one-loop corrections we decided to
work in the background field gauge \cite{Ab81,AbGrSc83} with the 
gauge parameter set to one. 
In this gauge the three-gluon
vertex is simplified which leads to a reduction 
of the number of terms encountered
in intermediate steps of the calculation. To reduce the one-loop tensor 
integrals to scalar one-loop integrals we used the Passarino-Veltman 
reduction procedure \cite{PaVe79}. From the diagrams shown in 
figure \ref{fig:loop-diagrams} one can immediately read off the scalar
box-integrals one encounters in the calculation. These integrals
are
\begin{equation}
  D_0^{d,l}
 =
 \frac{1}{i\pi^2}\int\!\!d^dl
 \frac{(2\pi\mu)^{(2\varepsilon)}}
 {(l^2+i\varepsilon)((l+k_g)^2+i\varepsilon)
   ((l+k_Q+k_g)^2-m^2+i\varepsilon)
   ((l-k_{\bar Q})^2-m^2+i\varepsilon)},\label{eq:lmaster}
\end{equation}
\begin{equation}
  D_0^{d,sl,1} 
 =
 \frac{1}{i\pi^2}\int\!\!d^dl
 \frac{(2\pi\mu)^{(2\varepsilon)}}
 {(l^2+i\varepsilon)((l+k_Q)^2-m^2+i\varepsilon)
   ((l+k_{Qg})^2-m^2+i\varepsilon)
   ((l-k_{\bar Q})^2-m^2+i\varepsilon)},
 \label{eq:sublmaster1}
\end{equation}
\begin{equation}
 \label{eq:sublmaster2}
 D_0^{d,sl,2}
   =\left.D_0^{d,sl,1}
 \right|_{k_Q\leftrightarrow k_{\bar Q}},
\end{equation}
with $k_{Qg}= k_Q+k_g$.
Note that only the real parts of the one-loop integrals contribute to
$h_6$ since we neglect
terms in the electroweak couplings $\sim {\rm Im} \chi$. For simplicity we
will leave out in the following the
prescript $\Re$ in front of the integrals.
All the lower-point integrals are defined with respect to the box-integrals.
We adopted  the notation from Passarino and Veltman \cite{PaVe79}
in which capital
letters $A$, $B$,$\ldots$ are used to define one-point, two-point,$\ldots$ 
integrals. The kinematics is specified by denoting as argument the 
propagators which are kept with respect to the defining box-integrals. 
For example $C_0^l(1,2,3)$ which originates from $D_0^l$ is given by
\begin{equation}
  C_0^l(1,2,3 )
  = 
  \frac{1}{i\pi^2}\int\!\!d^dl
  \frac{(2\pi\mu)^{(2\varepsilon)}}
  {(l^2+i\varepsilon)((l+k_g)^2+i\varepsilon)
    ((l+k_Q+k_g)^2-m^2+i\varepsilon)
   }.
\end{equation}
We have substituted the scalar box integrals in $d$ dimensions by
box integrals in $d+2$ dimensions using the relation
\begin{equation}
D_{27}^d = - {1\over 2\pi}D_0^{d+2},
\end{equation}
where $D_{27}^d$ is the coefficient of $g_{\mu\nu}$ in the decomposition
of the four-point tensor integral $D^d_{\mu\nu}$ 
(cf. eq. (F.3) in ref. \cite{PaVe79}),
which in turn can be expressed as a linear combination of the scalar box
integral $D_0^d$ and scalar triangle integrals $C_0$ in $d$ dimensions.
Since the box integrals are finite in 6 dimensions, all IR singularities
reside in triangle integrals after this substitution. As a gain, 
the coefficients of the triangle integrals become much simpler. 
This is clear, since
the coefficient that multiplies 
the sum of all IR singular terms has to be proportional to the Born result.
    
We may group the diagrams shown in figure \ref{fig:loop-diagrams}
according to their colour structure in terms of the colour matrices $T_a$ of
the SU(N) gauge group. In particular, we distinguish between contributions
which are dominant in the number of colours (`leading-colour')
and contributions which are suppressed in the number of colours 
(`subleading-colour'). The different colour structures are separately 
gauge independent.
The diagrams a, b, c are  proportional to $N (T_a)_{c_Qc_\Qb}$ 
where $a$ is the colour index of the gluon
and $c_Q$  $(c_\Qb)$ the colour index of the (anti-) quark.
The diagrams
h, j, k, l are proportional to $1/N (T_a)_{c_Qc_\Qb}$. They are purely
sub-leading in the number of colours.
The diagrams d, e, f, g are proportional to $(N-1/N)(T_a)_{c_Qc_\Qb}$.
These diagrams contribute to the leading as well as to the subleading
parts. 

To organize the next-to-leading order virtual corrections
$h_6^{\st{virt., NLO}}$ for $h_6$
it is convenient to separate them into different contributions. First
we split the result into a contribution from the loop diagrams 
(figure \ref{fig:loop-diagrams}) and
a contribution from the renormalization procedure $h_6^{\st{virt., ren.}}$.
The contribution from the loop-diagrams is further decomposed into 
a UV divergent ($h_6^{\st{virt., UV div.}}$), 
a IR divergent ($h_6^{\st{virt., IR div.}}$) and a finite part 
($h_6^{\st{virt., fin.}}$).
We thus arrive at the following decomposition:
\begin{equation}
  h_6^{\st{virt., NLO}}= 
   h_6^{\st{virt., UV div.}} 
  + h_6^{\st{virt., IR div.}}   
  + h_6^{\st{virt., fin.}}
  + h_6^{\st{virt., ren.}}.
\end{equation}
We define the singular contribution by taking only the pole-part
of the loop-integrals but keeping the $\e$-dependence in the prefactors.

\subsection{UV singularities}
All the UV singularities in a one-loop
calculation appear in the scalar one-  and two-point integrals. 
Defining the finite contributions $\overline{A}(m^2)$, $\overline{B}_0(i,j)$ 
of these integrals by
\begin{eqnarray}
  A(m^2) &=& \rg  \L{m^2\over 4\pi\mu^2}\R^{-\e} m^2 {1\over \e}
  + \overline{A}(m^2),\nn\\
  B_0(i,j) &=&
  \rg  \L{m^2\over 4\pi\mu^2}\R^{-\e} {1\over \e}
  + \overline{B}_0(i,j), 
\end{eqnarray}
where $\rg$ is the usual one-loop factor:
\begin{equation}
  \rg = {\Gamma(1+\e)\*\Gamma^2(1-\e)\over \Gamma(1-2\e)},
\end{equation}
the contribution from the pole-part of these integrals is given by
\begin{eqnarray}
  \label{eq:UVsingular}
  h_6^{\st{virt., UV div.}}
  &=& {\rg {1\over \e} \L{m^2\over 4\pi\mu^2}\R^{-\e}}
  \bigg\{
  {\as\over 2\*\pi} \*C_F\* \h6born
  +24 \*\as^2\*(N^2-1)\*C_F \* \deltah6\nn\\
  &+&8\*\as^2\* (N^2-1)\*C_F\* B\*\e\*\bigg( 
  1+x-2\*\xb+(1-\xb)\*{\xb\over x}\nn\\
  &+&z\*\bigg[2\*B\*
  ((\xb^2-2\*x)\*(1-x)+(2-3\*x)\*(1-\xb))
  +g_1\*{x^2-4\*z\over x\*(1-x)}
  +4\*z\*B\*(2\*{(1-\xb)^2\over 1-x}+\xg)\bigg]
  \bigg)\nn\\
  &+& 16\*\as^2\*C_F \*B\*\e\*(
  {1\over \xg}\*(x\*\xb-4\*\xb+2+\xb^2)
  -2\*z\*\xg\*B\*(x-\xb)
  ) + O(\e^2)\bigg\}
\end{eqnarray}
with
\begin{eqnarray}
  \deltah6 &=& B^2\*
  \bigg(\bigg[-2\*{(x^2-3\*x-\xb^2\*x+3\*x\*\xb+2\*\xb^2-4\*\xb+2)}
  -{\xg\*(\xb^2+x\*\xb-4\*\xb+2)}\*\e
  \bigg]\*z\nn\\
  &&-4\*{1\over 1-x}\*(-5\*\xb-3\*x+2\*\xb^2+4+x^2+x\*\xb)\*z^2
  \bigg),
\end{eqnarray}
and
\begin{equation}
  \label{eq:g_1def}
  g_1 = -{(1-x)\*(x-2\*\xb)\over (x^2-4\*z)^2}
  \*(x\*(\xb+x)-2\*(1-\xg))
  .
\end{equation}
The finite contributions involving $\overline{A}$, $\overline{B}_0$
are given in section \ref{sec:finite-oneloop}. The cancellation
of the UV singularities by the renormalization procedure will
be discussed in sec. \ref{sec:renormalization}.
 
\subsection{Soft and collinear singularities}
\label{sec:IR-divergencies}
To start with let us consider the soft and collinear singularities 
appearing in the leading-colour contribution.
As mentioned earlier we use the box-integral in $6-2\e$ dimensions 
rather than in $4-2\e$ dimensions. The one- and two-point 
integrals encountered in this calculation and
the box integrals in $6-2\e$ dimension do not contain IR singularities.
As a consequence all IR singularities appear in the triangle integrals.
Using the results for the integrals given in ref. \cite{BrUw97} 
we see that only $C_0^l(1,2,3)$ and $C_0^l(1,2,4)$ are IR divergent. 
The contribution from the two integrals is  given by
\begin{eqnarray}
  \left.h_6^{\st{virt., IR div.}}\right|_{\st{lead. colour}} 
  &=& -{\as\over 2\*\pi} \* {N}\*\h6born\* 
  \left\{
    (1-\xb)\,\*s\,\, \* C_0^l(1,2,3) + (1-x)\,\*s\,\,\*C_0^l(1,2,4)
  \right\}.
\end{eqnarray}
Using ref. \cite{BrUw97}
\begin{eqnarray}
  &&(1-\xb)\,\*s\,\, \* C_0^l(1,2,3)\nn\\
  &=&\rg \*
  \L {4\*\pi\*\mu^2  \over m^2 } \R ^{\e}
  \*\Bigg\{
  {1 \over 2} \*{1\over \e^2}
  - {1\over \e}\* \ln({t_{Qg}\over m^2})
  +\ln({t_{Qg} \over m^2})^2
  -{7\over 12}\*\pi^2-\Li2({{1\over \beta}})
  -{1\over 2}\*\ln(\beta)^2\Bigg\}+ O(\e) \nn\\
  &\equiv& 
  \rg \*
  \L {4\*\pi\*\mu^2  \over m^2 } \R ^{\e}
  \*\Bigg\{
  {1 \over 2} \*{1\over \e^2}
  - {1\over \e}\* \ln({t_{Qg}\over m^2})\Bigg\}
  +(1-\xb)\,\*s\,\, \*\overline{C}_0^l(1,2,3)
\end{eqnarray}
with 
\begin{equation}
  \beta = {1-\xb+z\over z}, \quad t_{ij} = 2k_i\cdot k_j,
\end{equation}
and $C_0^l(1,2,3)\stackrel{x\leftrightarrow \xb}{\longrightarrow}C_0^l(1,2,4)$
we obtain for the singular part
\begin{eqnarray}
   \left.h_6^{\st{virt., IR div.}}\right|_{\st{lead. colour}} 
   &=& -{\as\over 2\*\pi} \* \rg \*
  \L {4\*\pi\*\mu^2  \over m^2 } \R ^{\e}
  \*{N}\*\h6born\* 
  \left\{
    {1\over \e^2}
    - {1\over \e}\* \ln({t_{Qg}\over m^2})
    - {1\over \e}\* \ln({t_{\bar Qg}\over m^2})
  \right\}.  
  \label{eq:IR-div-LC}
\end{eqnarray}

In the sub-leading colour contribution only $C_0^{sl,1}(1,2,4)$ contains
singularities. The contribution from $C_0^{sl,1}(1,2,4)$ is given by
\begin{equation}
  \left.h_6^{\st{virt., IR div.}}\right|_{\st{subl. colour}} 
  = {\as\over2\*\pi }\*{1\over N}
  \*\h6born\*(x+\xb-2\*z-1)\,\*s\,\,\*C_0^{sl,1}(1,2,4).
\end{equation}
Using ref. \cite{BrUw97}
\begin{eqnarray}
  &&(x+\xb-2\*z-1)\,\*s\,\,\*C_0^{sl,1}(1,2,4) \nn\\
  &=& \rg
  \*\L {4\*\pi\*\mu^2\over m^2}\R^{\e}
  \*{1+\omega^2\over 1-\omega^2}
  \*\Bigg\{
  {1\over \e}\* \ln(\omega)  + {1\over 2}\* \ln^2(\omega)
  +2\*\,\Li2(1-\omega) -\pi^2\Bigg\}
  + O(\e)\nn\\
  &\equiv& \rg
  \*\L {4\*\pi\*\mu^2\over m^2}\R^{\e}
  \*{1+\omega^2\over 1-\omega^2}
  {1\over \e}\*  \ln(\omega)
  + (x+\xb-2\*z-1)\,\*s\,\,\*\overline{C}_0^{sl,1}(1,2,4)\nn\\
\end{eqnarray}
with
\begin{equation}\label{eq:omega}
  \omega = {1-\sqrt{1-{4\*z\over x+\xb-1}}\over
1+\sqrt{1-{4\*z\over x+\xb-1}}}
\end{equation}
we obtain 
\begin{eqnarray}
   \left.h_6^{\st{virt., IR div.}}\right|_{\st{subl. colour}} 
   ={\as\over2\*\pi }\*{1\over N}
  \*\h6born\* \rg
  \*\L {4\*\pi\*\mu^2\over m^2}\R^{\e} {1\over \e}
\*{1+\omega^2\over 1-\omega^2}\* \ln(\omega)
  \label{eq:IR-div-SC}
\end{eqnarray}
for the singular part.
There are additional IR singular contributions from
the counterterm diagrams. They will be discussed in section 
\ref{sec:renormalization}. Furthermore we note that $\h6born$ in the
equations above is the result in $d$ dimensions.
This follows immediately from the general properties of soft and collinear
limits in QCD. This is the reason why no
`$\gamma_5$-problem' arises from the IR singularities: Given the fact
that the IR singularities from the real corrections are also proportional
to the Born result in $d$ dimensions, finite contributions which would be
sensitive to the $\gamma_5$-prescription cancel together with the
IR singularities, as long as one uses the same $\gamma_5$-prescription for 
the virtual and the real corrections.

\subsection{Finite contributions}
\label{sec:finite-oneloop}
Given the divergent contributions shown in \eq{eq:UVsingular}, 
\eq{eq:IR-div-LC} and \eq{eq:IR-div-SC} we obtain
the following result for the finite parts:
\begin{equation}
  \label{eq:h6NLOfin}
  h_6^{\st{virt., fin.}} 
  = 
  \as^2 \,C_F N^2  \left.h_6^{\st{virt., fin.}}\right|_{\st{lead. colour}} 
  +\,\, \as^2\, C_F \left.h_6^{\st{virt., fin.}}\right|_{\st{subl. colour}}, 
\end{equation}
with
\begin{eqnarray}
\label{eq:lc-finite}
&&{1\over 4\* B}
\left.h_6^{\st{virt., fin.}}\right|_{\st{lead. colour}}=\nn\\
&&
-8\*\bigg\{-x\*(1-\xg)
+z\*B\*[x^3+5\*x^2\*\xb-6\*x^2+10\*x+5\*x\*\xb^2-16
\*x\*\xb+12\*\xb-10\*\xb^2-4+3\*\xb^3]\nn\\
&&-4\*B\*\xg\*z^2\*\bigg[ {(1-\xb)^2\over 1-x}+\xg\bigg] \bigg\}\* 
s\* {1 \over \pi} \* D_0^{d=6,l}
\nn\\
&&
+\bigg\{2\*B\*\bigg[1+4\*\xb\*(1-\xb)
+x\*(\xb-2)\*\xb+2\*(1-\xb)^2\*\xb\*{1\over x^2}
-(1-\xb)\*(1+2\*\xb\*(1-\xb))\*{1\over x}\bigg]\nn\\
&&+z\*\bigg[-8\*B\*(2\*{(1-\xb)^2\over 1-x}+\xg)
+{x^2\*\xb+8\*\xb-7\*x\*\xb-6\*x^2+24\*x-22
\over (1-x+z)\*(2-x)\*(1-x)}-{\xb^2+\xb-4\over (1-\xb+z)\*(1-\xb)}\nn\\
&&+8\*g_1\*{x^2-4\*z\over (2-x)\*(1-x)\*x^2}
\bigg]\bigg\}\*{1\over s}\*\overline{A}(zs)
\nn\\
&&
+\bigg\{6-10\*x-6\*(1-x)\*{1\over \xb}
+z\*\bigg[1-\xb+4\*x+g_2\*(1-{2\over \xb})
-4\*{1+x-4\*z\over 1-\xb}
-{\xb^2+\xb-4\over 1-\xb+z}\bigg]\bigg\}\*\overline{B}_0^{l}(1,3)
\nn\\
&&+\bigg\{16\*x
-z\*\bigg[g_2+3\*g_1\*x
+B\*\bigg(5\*x^2\*\xb+11\*x^2-55\*x+52\*x\*\xb
-13\*x\*\xb^2+64\*(1-\xb)-33\*\xb\*(1-\xb)\bigg)\nn\\
&&-\bigg(4\*x^3-x^2\*(2+3\*\xb)+x\*(4-\xb-4\*\xb^2)
+2-10\*\xb+14\*\xb^2
-4\*{1-\xb\over 1-x}
 \bigg)\*{1\over x^2-4\*z}\bigg]\nn\\
&&-32\*B\*z^2\*\bigg[1-x+{(1-\xb)^2\over 1-x}\bigg]\bigg\}\*
\overline{B}_0^{l}(2,3)
\nn\\
&&
-\bigg\{-2\*\xb^2\*{1\over x}-2\*\xb\*(1-\xg)\*{1\over x^2}
+z\*\bigg[-3+3\*g_1\*(2-x)-x+4\*\xb-{8\*\xb\over 1-x}\nn\\
&&+\bigg(10+2\*\xb+10\*\xb^2-x^2\*(4+3\*\xb)-x\*(2+\xb+4\*\xb^2)+4\*x^3
-8\*{1-\xb\over 2-x}+4\*{1-\xb\over 1-x}
-8\*(1-\xg)\*{\xb\over x^2}
-8\*{\xb^2\over x}
\nn\\
&&\bigg)\*
{1\over x^2-4\*z}
-{(x^2\*\xb+8\*\xb-7\*x\*\xb-6\*x^2+24\*x-22)\over (2-x)\*(1-x+z)}
\bigg]+{16\*z^2\over 1-x}\bigg\}\*\overline{B}_0^{l}(2,4)
\nn\\
&&
+2\*\bigg\{-x-3\*(1-\xg)\*{1\over \xb}+(1-\xb)\*\xb\*{1\over x^2}
-\xb\*(1+\xb)\*{1\over x}
+z\*\bigg[3\*g_1+g_2\*{1\over \xb}+{8\*\xg\over 1-x}\nn\\
&&+(2+6\*\xb-2\*x^2-x\*(3-\xb)+4\*{1-\xb\over 1-x}
+4\*(1-\xb)\*\xb\*{1\over x^2}-4\*\xb\*(1+\xb)\*{1\over x}
)\*{1\over x^2-4\*z}\bigg]\bigg\}\*\overline{B}_0^{l}(3,4)
\nn\\
&&
-8\*(1-\xb)\*\bigg\{x-{2\*\xg\*z\over 1-x}\bigg\}
\*s\,\*\overline{C}_0^{l}(1,2,3)
-8
\*(1-x)\*\bigg\{x-{2\*\xg\*z\over 1-x }\bigg\}\*s\,\*\overline{C}_0^{l}(1,2,4)
\nn\\
&&
+2\*z\*(1-\xb)\*\bigg\{{1\over 2}\*g_2
+7 +2\*x
-{1\over 2}\*(1+\xb)\*\bigg[1
+{8\over 1-x}\bigg]
+8\*z\*\bigg[{1\over 1-\xb}+{2\*\xg\over (1-x)^2}\bigg]
\bigg\}\*s\,\*C_0^{l}(1,3,4)
\nn\\
&&
-2\*z\*\bigg\{-{3\over 2}\*g_1\*(1-x)\*x-{3\over 2}+3\*x+{1\over 2}
\*x^2+4\*{1-x^2\over 1-\xb}-2\*x\*\xb
-8\*z\*\bigg[1+2\*{1-x\over 1-\xb}\bigg]\nn\\
&&-{1\over 2}\*\bigg[18\*x\*\xb^2-14\*\xb^2-4\*\xb^2\*x^2+2\*x^2\*\xb+6\*\xb
-9\*x\*\xb-3\*x^3\*\xb+4\*x^4\nn\\
&&
+2\*(1-x)\*(1+3\*x^2)\bigg]\*{1\over x^2-4\*z}
\bigg\}\*s\,\*C_0^{l}(2,3,4)
\nn\\
\end{eqnarray}
and
\begin{eqnarray}
\label{eq:sc-finite}
&&{1\over 4\*B}\left.h_6^{\st{virt., fin.}}\right|_{\st{subl. colour}} =\nn\\
&&
-2\*\bigg\{ 2\*(1-\xb) + z\*\bigg[\xb - 2\* {1-2\*x\over 1-\xb}
-4\*{z\*\xg\over (1-\xb)^2}\bigg]\bigg\}\*\overline{B}_0^{l}(1,3)\nn\\
&&
+
4\*z\*\bigg\{-B\*(2\*x^2+3\*x\*\xb-5\*x-x\*\xb^2+3-2\*\xb)
+2\*z\*B^2\*(x-\xb)\*\xg^2-{(1-x)\*(x-2\*\xb)\over x^2-4\*z}\nn\\
&&+{(\xb-2)\*\xg\over 1-\xg-4\*z}\bigg\}\*\overline{B}_0^{l}(2,3)\nn\\
&&+2\*\bigg\{2\*\xb\*{1-x\over x}
+z\*\bigg[4-\xb+2\*{\xb\over 1-x}
-4\*{z\*\xg\over (1-x)^2}
-2\*\bigg\{
  2\*(1+\xb)-x-4\*{\xb\over x}
  \bigg\}\*{1-x\over x^2-4\*z}
\bigg]\bigg\}\*\overline{B}_0^{l}(2,4)\nn\\ 
&&+4\*\bigg\{\bigg[x^2-\xb\*(\xb-2)\*(2+(\xb-2)\*{1\over x})
-2\*(1-\xg)\bigg]
\*{1\over \xg^2}
+2\*z\*{(x-2\*\xb)\*(1-x)\over x\*(x^2-4\*z)}
\bigg\}\*\overline{B}_0^{l}(3,4)\nn\\
&&-4\*\bigg\{
-\bigg[\xb\*(\xb+x)-2\*(1-\xg)\bigg]\*{1\over \xg^2}
+z\*{(\xb-2)\*\xg\over 1-\xg-4\*z}\bigg\}\*\overline{B}_0^{sl,1}(2,4)\nn\\
&&-z\*(1-\xb)\*\bigg\{g_2+1
+4\*x
+3\*\xb+2\*{\xb\*(1-x)\over 1-\xg-4\*z}\bigg\}
\*s\,\*C_0^{l}(1,3,4)\nn\\
&&+2\*z\*\bigg\{-\xb\*(1+\xb)+4\*(1-\xg)+4\*z\*{\xb-x\over 1-\xb}
-{\xb\over (1-\xg-4\*z)\*B}\bigg\}\*s\,\*C_0^{sl,1}(1,2,3)\nn\\
&& +8\*(x+\xb-2\*z-1)\* \bigg\{ x-{2\*\xg\*z\over 1-x}\bigg\}
\*s\,\*\overline{C}_0^{sl,1}(1,2,4)\nn\\
&&+2
\*z\*\bigg\{
2\*
\bigg[
3\*(1-x)-\xb\*(1-x)+{(\xb-2)\*(1-\xb)\over 1-x}
-2\*{(1-x^2)\over (1-\xb)}
+B\*{\xb\*(1-\xb)^2\over 1-\xg}\bigg]
\nn\\
&&+8\*z\*B\*
\bigg[\xg\*(1+\xg)-(1-x)\*x-{1-\xb\over 1-\xg}\bigg]
+2\*{(x-2\*\xb)\*(1-x)^2\over x^2-4\*z}\nn\\
&&
-\bigg[x\*(4-\xb)-\xb-2\*{\xb\*(1-\xb)\over 1-x}\bigg]
\*{1-\xb\over 1-\xg-4\*z}\bigg\}\*s\,\*C_0^{l}(2,3,4)\nn\\
&&
+2\*z\*\bigg\{
2\*
(1-\xb)\*\bigg[
\xg\*(1 -2\*{1-\xg\over (1-\xb)^2})
+{4-3\*\xb\over 1-x}+{1\over \xb\*(1-x)}
+{1\over (1-\xg)\*\xb}
+{(1-\xb)^2\over (1-x)^2}
+4\*{1\over \xg}\bigg]
\nn\\
&&
-8\*z\*B\*\bigg[1+\xg-x\*{1-x\over 1-\xb}
 +{(1-\xb)^2\over 1-x}
\bigg]\*{\xg\*(1-\xb)\over 1-\xg}\nn\\
&&
-\bigg[4 -\xb - \xb\*{1-x\over 1 -\xb} +2\*{\xb-2\over 1-x}
+2\*\xb\*{1-\xb\over (1-x)^2}\bigg]\*{\xg\*(1 -\xb)\over 1-\xg-4\*z}
\bigg\}\*s\,\*C_0^{sl,1}(2,3,4)\nn\\
&&-2\*z\*\bigg\{
\bigg[-6+5\*\xb +{x\*(2+\xg)\*(2-\xb)\over 1-\xb}
+2\*\xb\*{1-\xb\over 1-x}\bigg]\*{1-\xb\over 1-\xg}\nn\\
&&+{4\*z\over 1-\xg}\*\bigg[x-6+4\*\xb-{(1-\xb)\*(2-3\*\xb)\over 1-x}\bigg]
\nn\\
&&
+\bigg[x\*(4-\xb)-\xb-2\*\xb\*{1-\xb\over 1-x}\bigg]
\*{1-\xb\over 1-\xg-4\*z}
\bigg\}\*s\,\*C_0^{sl,2}(1,3,4)\nn\\
&&+2\*\bigg\{
2\*(x-\xg)\*(1-\xg)\nn\\
&&+z\*\bigg[14\*(1-x)-2\*\xb\*(4-\xg)
-4\*x\*{1-x\over 1-\xb}
+8\*z\*{\xg\over 1-\xb}-{\xb\*(1-x)\*(x-\xb)\over 1-\xg-4\*z}
\bigg]\bigg\}\*{s\over \pi}\*
D_0^{d=6,sl,1}\nn\\
&&+2\*\bigg\{
2\*(\xb-x)\*(1-\xb)-2\*x+z\*\bigg[-2\*\bigg\{
13-8\*\xb-2\*x-\xg^2
-\xb\*{(1-\xb)^2\over (1-x)^2}\nn\\
&&
-2\*{\xb\*(4-3\*\xb)\over 1-x}
-{x\*(3+x)\over 1-\xb}
+4\*z\*\xg\*\bigg\{(2-\xb)\*B+{1-\xb\over (1-x)^2}\bigg\}\bigg\}\*{1-\xb\over 1-\xg}\nn\\
&& 
-\bigg\{4-6\*\xb+x\*(4-\xb)+\xb^2-4\*{1-\xb\over 1-x}
+2\*\xb\*{(1-\xb)^2\over (1-x)^2}\bigg\}
\*{1-\xb\over 1-\xg-4\*z}\bigg]\bigg\}\*{s\over \pi}\*D_0^{d=6,sl,2}
\nn\\
&& - {1\over 4\*B} \* \bigg\{ a \* \overline{A}(zs)
   +b_{13}\* \overline{B}_0^{l}(1,3)
   +b_{23}\* \overline{B}_0^{l}(2,3)
   +b_{24}\* \overline{B}_0^{l}(2,4)
   +b_{34}\* \overline{B}_0^{l}(3,4)
   +c_{234}\* C_0^{l}(2,3,4)\bigg\}
,\nn\\
\end{eqnarray}
with $g_1$ defined in \eq{eq:g_1def},
\begin{eqnarray}
  g_2 &=& {1\over \xb^2-4\*z} \*
  (12\*(1-x) - 3\*\xb^3-6\*x\*\xb^2+13\*\xb^2-20\*\xb+16\*x\*\xb)
  ,
\end{eqnarray} 
and $B$ defined in \eq{eq:Bdef}.
The coefficients $a,\,\, b_{13},\ldots$ of the loop-integrals in the 
last line of 
\eq{eq:sc-finite}
are the same as the coefficients of the corresponding integrals in 
\eq{eq:lc-finite}.

\subsection{Renormalization}
\label{sec:renormalization}
\begin{figure}[htbp]
  \begin{center}
    \leavevmode
    \centerline{\epsfxsize9cm\epsfbox{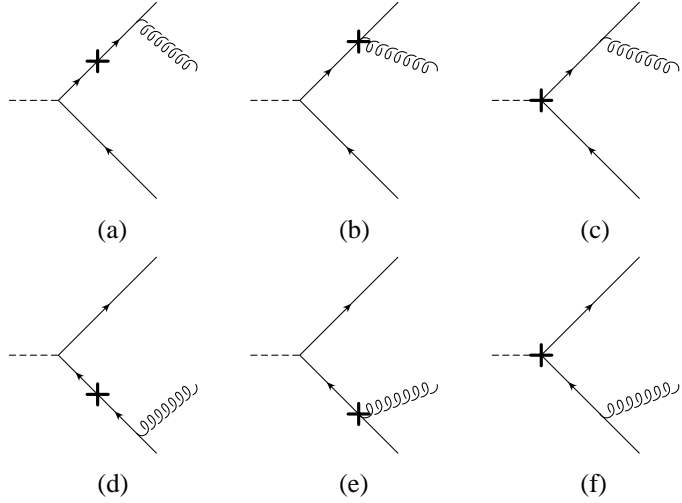}}
    \caption{Counterterm diagrams to $Z\to Q\Qb g$.}
    \label{fig:ct-diagrams}
  \end{center}
\end{figure}

In figure \ref{fig:ct-diagrams} we show the Feynman diagrams which 
arise from ${\cal L}_{ct}(\Psi_R,A_R, m_R,g_R)$ in 
\eq{eq:ren-pert}.
The `mass counterterm' shown in the diagrams a, d is given by
\begin{equation}
  i((Z_\Psi-1)\ksl -(Z_0-1)m_R)
  \equiv
  i(\delta Z_\Psi \ksl - \delta Z_0 m_R)
  =  i \delta Z_\Psi (\ksl-m_R) +i (\delta Z_\Psi- \delta Z_0 )m_R,
\end{equation}
with the momentum counted in the direction of the fermion flow.
The first term will result in contributions proportional
to the tree amplitude, while the second term will generate 
a new structure which is not proportional to
the tree amplitude. This term should not contain infrared divergences.
Inserting the explicit results for
$\delta Z_\Psi$, and $\delta Z_0$ it can be easily checked that
this term is indeed IR finite. The other contribution from the
remaining counterterms (diagrams b, c, e, f) are proportional to the 
tree amplitude.
Using the relation $Z_g = (Z_A)^{-1/2}$ \cite{Ab81} valid in the
background field gauge as long as one uses the MS or \msbar\ scheme
for the gluon field and coupling,
we find that the gluon renormalization enters only through
the residue ${\cal Z}_A$ of the renormalized gluon propagator.
The contribution from the counterterms is thus given by
\begin{equation}
  (2\delta Z_\Psi -\delta Z_\Psi) {\cal T}^{\st{tree}}
  + \tilde {\cal T}^{\st{ct, a+d}}
  =  \delta Z_\Psi {\cal T}^{\st{tree}}
  + \tilde {\cal T}^{\st{ct, a+d}}
\end{equation}
with $\tilde {\cal T}^{\st{ct, a+d}}$ the contribution from
the insertion of $i (\delta Z_\Psi- \delta Z_0 )m_R$.
For the squared amplitude we get
\begin{eqnarray}
  &&  
  2\delta Z_\Psi \h6born
  - 32\*\pi \as(N^2-1)(\delta Z_\Psi- \delta Z_0 )\*
  \deltah6.   
\end{eqnarray}
In the on-shell scheme we have
\begin{eqnarray}
  \delta Z_\Psi^{\st{on}} &=& 
  {\as\over 4\*\pi}\*\rg\*C_F 
  \L{m^2\over 4\*\pi\*\mu^2}\R ^{-\e} \* \left\{
    -{1\over \e}
    +{2\over \e_{\st{IR}}} 
    - 4 \right\} + O(\e)
\end{eqnarray}
and
\begin{eqnarray}
  \delta Z_\Psi^{\st{on}}- \delta Z_0^{\st{on}} 
  &=&
  {\as\over 4\*\pi}\*C_F \rg \L {m^2\over 4\pi\mu^2}\R ^{-\e}
  \left\{  {3\over \e} + 4 \right\}+ O(\e).
\end{eqnarray}
To distinguish between infrared and ultraviolet singularities we have 
introduced  $\e_{\st{IR}}$ by $d=4+2\e_{\st{IR}}$. This allows us to check 
separately the cancellation of the UV and IR singularities.
To finish the renormalization procedure we must include the
contribution from the residue of the gluon propagator ${\cal Z}_A^\msbar$: 
\begin{equation}
  ({\cal Z}_A^\msbar)^{{1\over 2}} {\cal T}^{\st{tree}}
\end{equation}
which amounts to 
\begin{equation}
  2 (({\cal Z}_A^\msbar)^{1/2}-1) |{\cal T}^{\st{tree}}|^2
\end{equation}
at the level of the squared matrix element.
(In the on-shell scheme the residue ${\cal Z}_\Psi$ 
of the renormalized quark propagator  
is one by definition.) 
We have 
\begin{eqnarray}
  {\cal Z}_A^\msbar
  &=& 
  1-{\alpha_s\over 4\*\pi}\*{1\over 3}{\L 4\pi\R^\e\over \Gamma(1-\e)}
  \Big[{1\over \e_{\st{IR}}}
  \*(2\* n_f^l -11\* N) 
  -2\* \sum_i \ln{m_i^2\over \mu^2}\Big] +O(\e) \equiv 1 
  -\delta {\cal Z}_A^\msbar,
\end{eqnarray}
where $n_f^l$ is the number of massless flavours and the $m_i$ are the 
masses of the massive quarks, including the flavour of \eq{eq:QQbg},
that contribute to the gluon self energy.
We can now write down the entire contribution from UV renormalization:
\begin{eqnarray}
  \left.h_6^{\st{virt., ren.}}\right|_{\st{UV}}
  &=&(2\delta Z_\Psi^{\st{on}} - \delta {\cal Z}_A^\msbar )\h6born
  - 32\*\pi \as(N^2-1)(\delta Z_\Psi^{\st{on}}- \delta Z_0^{\st{on}} )\*
  \deltah6 \nn\\
  &=&
  -{\as\over 2\*\pi}\* \rg\*C_F 
  \L{m^2\over 4\pi\mu^2}\R ^{-\e} \* 
   {1\over \e}\h6born
  - 24\as^2   (N^2-1) C_F
  \rg \L {m^2\over 4\pi\mu^2}\R ^{-\e} {1\over \e} 
  \*
  \deltah6\nn\\
  &+&{\as\over 4\*\pi}\*\rg\*  \L{m^2\over 4\pi\mu^2}\R ^{-\e} 
  {1\over \e_{\st{IR}}}\left\{ 4  C_F 
  - {1\over 3} 
  \*\left(2\* n_f^l -11\* N\right) 
  \right\}\h6born \nn\\
  &+&{\as\over 4\*\pi}\*  
  \left\{ {1\over 3}\*(2\* n_f -11\* N)\ln({m^2\over \mu^2})
  +{2\over 3}\*\sum_i \ln{m_i^2\over m^2} - 8  C_F\right\}  \h6born
  - 32\as^2   (N^2-1) C_F
  \deltah6.\nn\\ 
  \label{eq:ren-contr}
\end{eqnarray}
Comparing this result with \eq{eq:UVsingular} it follows immediately that all
the 
UV singularities cancel. In addition it can be seen 
that the factor $\mu^{2\e}$ in front of the UV singularities cancels together
with the singularities. As we will show in section \ref{sec:IR-cancellation}
the same is true for the $\mu$-dependence in front of the IR singularities.
The remaining  $\mu$-dependence is thus given by
\begin{equation}
  {\as\over 4\*\pi}\*  
  \beta_0 \ln({\mu^2\over m^2 }) \h6born
  \quad
  \mbox{with}
  \quad \beta_0= {1\over 3}\*(11\* N-2\* n_f )
\end{equation}
as it should be according to the renormalization group equation. 
However this is only true as long as we renormalize the mass parameter
in the on-shell scheme. 

We close this subsection with some remarks concerning the
treatment of $\gamma_5$. Since we perform the computation of $h_6$ in $d$
dimensions, we have to give a prescription how to treat $\gamma_5$.
We use the following substitution in the axial
vector current (remember that $\mu$ is a 4-dimensional index) 
\cite{tHoVe72,La93}:
\begin{equation}
     \label{eq:g5-larin}
      \gamma_\mu \gamma_5\rightarrow Z_5^{ns}\frac{i}{3!}\,
      \varepsilon_{\mu\beta\gamma\delta}\,
      \gamma^\beta\gamma^\gamma\gamma^\delta,
      \qquad\mbox{with}\quad Z_5^{ns} = 1 -\frac{\as}{\pi} C_F +
\order(\as^2).
\end{equation}
The finite `counterterm' $(Z_5^{ns}-1)$ induces a contribution
\begin{equation}
  \label{eq:gamma5_ct}
  \left.h_6^{\st{virt., ren.}}\right|_{{\gamma_5}} =
  -\frac{\as}{\pi} C_F \h6born.
\end{equation}
This term is necessary to restore the chiral Ward identities,
which 
can be checked in the present case by verifying the following
relation:
\begin{equation}
  \label{eq:ward}
  h_6^{VA}(m=0)=h_6^{AV}(m=0).
\end{equation}  
Here, the index ${VA}({AV})$
denotes the contribution to $h_6$ from the interference of the Born
(1-loop) vector current with the 1-loop (Born) axial vector current.
We have checked that after inclusion of the term 
given in \eq{eq:gamma5_ct}
the relation shown in \eq{eq:ward} is indeed satisfied.

\subsection{Checks of the calculation}
The finite parts of the loop contributions
to the virtual corrections are given by the rather
complicated expressions \eq{eq:lc-finite} and \eq{eq:sc-finite}. 
It is 
important to perform some checks to ensure their correctness. To this
end it is useful to realize that the decomposition in terms of
(coefficients)$\times$(one-loop integrals) is unique in the sense
 that in the massless case the 
coefficients of the loop-integrals can be uniquely reconstructed from
the coefficients of specific logarithms even after having inserted the 
analytic expressions for the loop-integrals.

Most of the coefficients are highly constrained due to the general
properties
of infrared finiteness and renormalizability. As already mentioned
above,
only the $A$ and $B_0$ integrals contain UV singularities. Therefore
the fact that all UV singularities cancel after inclusion of the
counterterms
is an excellent check of the coefficients of the one- and
two-point integrals, which are each quite involved but combine to give
the
rather simple expression \eq{eq:UVsingular}.
As discussed in section \ref{sec:IR-divergencies}, all IR singularities
reside in three-point integrals. They are given explicitly in 
\eq{eq:IR-div-LC} and \eq{eq:IR-div-SC}. Therefore the
coefficients
of the IR divergent $C_0$ integrals are constrained by the requirement
that the IR singularities in the virtual corrections cancel against the
IR singularities from the real emission processes. We will see in the
next section that this cancellation really takes place
(cf. \eq{eq:v-singular} and \eq{eq:r-singular}).
Note that only the coefficients of the 
IR finite $C_0$ integrals
are not tested by the consideration above. (The coefficients of the
box-integrals enter the coefficients of the divergent $C_0$ integrals
and are thus also tested.) 
Since {\it all} coefficients of the loop integrals are computed
simultaneously by one code, the above considerations provide already a
quite powerful check. 

We can, however, perform an additional test by comparing
our result in the limit $m\to 0$ with the known massless result
for $h_6$ given in references \cite{KoSc85,KoSc89},
and \cite{KoScKrLa86}.
In these references the loop-integrals are substituted by their analytic
expressions in terms of logarithms and dilogarithms. 
As mentioned above we can reconstruct the coefficients of the loop-integrals
from the coefficients of specific logarithms and dilogarithms. For
example in the massless case the dilogarithm appears only in the 
box-integrals. As a consequence the coefficient of the function
$r(x,y)$ defined in eq.~(4.11) of ref. \cite{KoSc89} must be 
proportional to the coefficients
of the box-integrals in our result, if we consider the massless case.
Note that the massless case cannot be derived by simply taking the 
massless limit, because the massless limit of the loop-integrals does
not exist in general. However we can still compare the coefficients
of the loop integrals with the massless result by setting $z=0$ in the
coefficients and checking which logarithm/dilogarithm in the results of ref. 
\cite{KoSc89} are generated from which loop-integrals if one calculates
the integrals for massless quarks.
We do not need to worry about the additional collinear singularities
which are present in the loop-integrals for massless quarks,
as long as we want to 
check only the coefficients of the loop-integrals.

In this way we can compare our result for example to the one 
given in eq.~(4.9) and eq.~(4.10) of ref. \cite{KoSc89}. 
The definition of $H_{6,7}$ in ref. \cite{KoSc89} differs by a
factor of 2 from ours due to a different convention in the definition of the
electroweak couplings. Taking this into account, we find complete
agreement of our result
with the coefficients of the $\ln^2$, $\ln$ and $r(x,y)$ functions
in eq.~(4.10) of ref. \cite{KoSc89}. We also can reproduce their rational
function,
but only up to a multiple of the Born result. This is not surprising,
since
the massless limit induces additional divergent and finite terms and
since we
use a different method to isolate the IR divergences in the real
emission processes. For the leading
contribution in the number of colours
we find agreement of our coefficients  with those of eq.~(4.9)   in
ref. \cite{KoSc89} up to an overall factor $(-1)$.
This is most probably caused by  a typographical error  
which is also present in ref. \cite{KoSc85},  but not in 
ref. \cite{KoScKrLa86} with which we fully agree.

\section{Singular contributions from real emission}
\label{sec:realcontributions}
In this section we discuss the cancellation of the soft and collinear 
singularities appearing in the loop corrections in particular 
\eq{eq:IR-div-LC}, \eq{eq:IR-div-SC} and \eq{eq:ren-contr}. To cancel them
we must include the singular contributions from the real emission
processes, in particular $\epem \to Q\Qb q\qb$ and $\epem \to Q\Qb gg$,
with $q$ denoting a light quark.
The reactions $\epem \to Q\Qb Q'\Qb'$ and $\epem \to Q\Qb Q\Qb$ do
not yield a singular contribution. As a consequence these 
contributions can be integrated numerically in 4 dimensions.
In $\epem \to Q\Qb Q'\Qb', Q\Qb Q\Qb$
the collinear singularities which would appear for massless quarks
are 
regulated by the quark mass and give rise to $\ln(m^2)$ terms. To a certain
extend these logarithms cancel in the sum of virtual and real contributions. 
On the other hand in tagged cross sections one encounters
in general the situation where uncancelled logarithms remain in the 
final result. 
For example in $\epem \to Q\Qb q \qb$ it can be easily seen that 
such logarithms will be produced when the heavy quark pair is produced
by a gluon emitted from the light quarks. Due to the fact that this
contribution is proportional to the couplings of the light quarks to the 
Z~boson it can never be cancelled by the virtual contributions.
In the presence of  several scales of different
orders of magnitude these 
logarithms may spoil the convergence of 
fixed order perturbation theory and indicate a sensitivity to long distance
phenomena. (Because of these logarithms it is for example
not longer possible to take
the massless limit!)  The uncancelled logarithms may need a treatment by
renormalization group techniques which goes beyond the fixed order
perturbation theory. In particular they should be absorbed into
the appropriate fragmentation functions.
These contributions can also be isolated by  experimental cuts 
\cite{Aleph98,Delphi99,SLD99}.
  
To cancel soft and collinear singularities several algorithms have been
developed in the past \cite{GiGl92,FrKuSi96,CaSe97}. 
In this work we use the \pss
method which was modified to include mass effects
in $e^+e^-\to 3$ jets in references \cite{BeBrUw97a,BrUw97} and 
generalized to treat massive quarks in arbitrary jet cross
sections in ref. \cite{KeLa99}.
Here  we restrict ourself to the singular parts from the
real emission processes. The finite contributions which can be
calculated by numerical integration of the existing leading order 
matrix elements will be discussed elsewhere. 
The idea of the \pss is to split the phase space of the real emission
contributions into regions where all the partons are hard and not 
collinear {\it (resolved regions)} and the remaining 
regions where two partons are collinear or
a parton is soft {\it (unresolved regions)}. 
The separation into resolved and unresolved regions can be performed 
by introducing appropriate Heaviside functions. In the unresolved
regions one can make use of the soft and collinear factorization
to approximate the matrix elements. Due to this simplifications it is
possible to perform the integrations analytically over the unresolved
regions.

To fix the notation and
to define as clearly as possible resolved and unresolved contributions
we outline in the next two sections the \pss method applied
to the problem at hand.

\subsection{Singular contribution from $Q\bar Q q\bar q$}
The matrix element for $e^+e^- \to Q(k_Q)\Qb(k_\Qb) q(k_q)\qb(k_\qb)$
is singular in the limit $k_q||k_\qb$. Defining $t_{ij} = 2k_i\cdot k_j$
we may introduce the following slicing:
\begin{equation}
  |\T(Q(k_Q)\Qb(k_\Qb) q(k_q)\qb(k_\qb))|^2
  = [\Theta(t_{q\qb}-\smin) + \Theta(\smin-t_{q\qb})]
  |\T(Q(k_Q)\Qb(k_\Qb) q(k_q)\qb(k_\qb))|^2,
  \label{eq:QQqqslicing}
\end{equation}
where $\smin$ denotes an arbitrary mass scale squared 
which separates between
collinear and resolved regions. The resolved contribution 
\begin{equation}
  d\sigma_R=\Theta(t_{Q\Qb}-\smin)|\T(Q(k_Q)\Qb(k_\Qb) q(k_q)\qb(k_\qb))|^2
  dR^d(k_Q,k_\Qb,k_q,k_\qb)
\end{equation}
is free of collinear singularities 
and can thus be integrated numerically in 4 dimensions.\\
Here $dR^d(k_Q,k_\Qb,k_q,k_\qb)$ denotes the four particle phase space
measure in $d$ dimensions:
\begin{equation}
  dR^d(k_1,\ldots,k_n) 
  = 
  {1\over {\cal N}}
  (2\pi)^d \delta^{(d)}(K-\sum_{i=1}^n k_i)
  \prod_{i=1}^n {d^{d-1} k_i\over (2\pi)^{d-1} 2 k_i^0}.
\end{equation}
(${\cal N}$ is a symmetry factor.)
For small $\smin$ the second term in \eq{eq:QQqqslicing} 
 can be approximated using
the factorization of amplitudes in the collinear limit:
\begin{eqnarray}
  &&\Theta(\smin-t_{q\qb}) |\T(Q(k_Q)\Qb(k_\Qb) q(k_q)\qb(k_\qb))|^2\nn\\
  &\stackrel{k_q||k_\qb}{\longrightarrow}&
  \Theta(\smin-t_{q\qb}) |\T(Q(k_Q)\Qb(k_\Qb) g(k_q+k_\qb))|^2
  g_s^2  {2\over t_{q\qb}} P_{q\qb \to g}(z)
\end{eqnarray}
where $P_{g\to q\qb}(z)$ is the Altarelli-Parisi 
kernel \cite{AlPa77} (in conventional dimensional regularization)
\begin{equation}
  P_{q\qb \to g}(z) = {1\over 2}{z^2+(1-z)^2-\e \over 1-\e}
\end{equation}
and $z$ denotes the momentum fraction in the collinear limit:
\begin{equation}
   k_q = z (k_q+k_\qb).
\end{equation}
Using in addition the factorization of the phase-space measure
in the collinear limit \cite{GiGl92}:
\begin{eqnarray}
  dR^d(k_Q,k_\Qb,k_q,k_\qb) 
  \stackrel{k_q||k_\qb}{\longrightarrow}
  dR^d(k_Q,k_\Qb,P)\dRcoll(k_q,k_\qb)
\end{eqnarray}
with 
\begin{equation}
  \dRcoll(k_q,k_\qb) = 
  {1\over16\*\pi^2}\*{\L{4\*\pi\mu^2}\R ^{\e}\over\Gamma(1-\e)}\*
  \,dt_{q\qb} dz [t_{q\qb} z (1-z)]^{-\e}
\end{equation}

the unresolved contribution to the cross section  after integration
over the collinear region is given by
\begin{eqnarray}
  d\sigma_{U}(Q(k_Q)\Qb(k_\Qb) q(k_q)\qb(k_\qb)) =
  d\sigma(Q(k_Q)\Qb(k_\Qb)g(k_g)) \cC_{q\qb}
  \label{eq:QQqq-singular}
\end{eqnarray}
with
\begin{eqnarray}
  \cC_{q\qb} 
  &=& 
  -{g_s^2\over 8\*\pi^2}\*{1\over\Gamma(1-\e)}
  \*\L{4\*\pi\mu^2\over \smin}\R ^{\e}
  {1\over \e}\int_0^1 dz [z (1-z)]^{-\e} 
  P_{q\qb \to g}(z)\nn\\
  &=& 
  -{\as\over2\*\pi}\*{1\over\Gamma(1-\e)}
  \*\L{4\*\pi\mu^2\over \smin}\R ^{\e}
  {1\over \e}\*
  {1-\e\over 3-2\*\e} \* {\Gamma^2(1-\e)\over \Gamma(2-2\*\e)}.
\end{eqnarray}
In \eq{eq:QQqq-singular} we have identified 
the sum of momenta $k_q+k_\qb$ with the momentum
$k_g$ of the outgoing `recombined' gluon.
In particular the singular contribution to $h_6$ is given by 
$\cC_{q\qb} \times \h6born.$ We get the same contribution for each massless
quark flavour. This results to an overall factor $n_f^l$ with $n_f^l$
the number of light flavours.

\subsection{Singular contribution from $Q\bar Q gg$}
The singular behavior of the matrix element for 
$\epem \to Q(k_Q)\Qb(k_\Qb)g(k_1)g(k_2)$ is far more complicated
than that discussed in the previous section. While in  
$\epem \to Q\Qb q\qb$ only collinear singularities appear we have now
both: soft and collinear singularities. In addition only the colour ordered
sub-amplitudes show a simple factorization in the soft limit. So we
may start with the colour decomposition of the amplitude which reads
\begin{equation}
  \T(Q(k_Q)\Qb(k_\Qb)g(k_1)g(k_2)) = 
  (T_{a_1} T_{a_2} )_{c_Qc_\Qb} S_1 + (T_{a_2} T_{a_1} )_{c_Qc_\Qb} S_2.
\end{equation}
Here $c_Q$ ($c_\Qb$) is the colour index of the (anti-) quark, and 
$a_1$, $a_2$ are the colour indices of the gluons. In our conventions
the colour matrices satisfy $\tr[T_a T_b] = \delta_{ab}$.
The amplitudes 
$S_1$, $S_2$ are the so called colour ordered sub-amplitudes.
The squared amplitude in terms of the  sub-amplitudes is
given by 
\begin{equation}
  \sum_{\st{colour}} |\T(Q(k_Q)\Qb(k_\Qb)g(k_1)g(k_2))|^2
  =
   (N^2-1)\* \L N\*(|S_1|^2+|S_2|^2)
  -{1\over N} |S_1+S_2|^2 \R .
\end{equation}
Note that the sub-amplitudes are gauge independent. We can thus study 
the soft and collinear limits independently for  $|S_1|^2$, $|S_2|^2$
and $|S_1+S_2|^2$. The contribution $|S_1+S_2|^2$ is essentially 
QED-like. It is therefore free of collinear singularities in the limit
that the two gluons become collinear.  
The amplitude $S_2$ can be obtained from $S_1$ by exchanging
the two gluon momenta. So let us start with the treatment of $S_1$.
The appropriate invariants for the \pss are $t_{Q1}$, $t_{12}$ and $t_{2\Qb}$.
We start with the following identity:
\begin{eqnarray}\label{eq:theta}
  1&=&[\Theta(t_{Q1}-\smin)+\Theta(\smin-t_{Q1})]
  \*[\Theta(t_{12}-\smin)+\Theta(\smin-t_{12})]\nn\\
  &&\times[\Theta(t_{2\Qb}-\smin)+\Theta(\smin-t_{2\Qb})]\nn\\
  &=&
  \Theta(t_{Q1}-\smin)\*\Theta(t_{12}-\smin)\*\Theta(t_{2\Qb}-\smin)\nn\\
  &+&\Theta(t_{Q1}-\smin)\*\Theta(\smin-t_{12})\*\Theta(t_{2\Qb}-\smin)
  \mbox{ 1,2 collinear} \nn\\
  &+&\Theta(t_{Q1}-\smin)\*\Theta(\smin-t_{12})\*\Theta(\smin-t_{2\Qb})
  \mbox{ 2 soft}\nn\\
  &+&\Theta(\smin-t_{Q1})\*\Theta(\smin-t_{12})\*\Theta(t_{2\Qb}-\smin)
  \mbox{ 1 soft}\nn\\
  &+&\Theta(t_{Q1}-\smin)\*\Theta(t_{12}-\smin)\*\Theta(\smin-t_{2\Qb})
  \mbox{ \Qb,2 collinear}\nn\\
  &+&\Theta(\smin-t_{Q1})\*\Theta(t_{12}-\smin)\*\Theta(t_{2\Qb}-\smin)
  \mbox{ Q,1 collinear}\nn\\
  &+&\Theta(\smin-t_{Q1})\*\Theta(t_{12}-\smin)\*\Theta(\smin-t_{2\Qb})
  \mbox{ Q,1 collinear and \Qb,2 collinear}\nn\\
  &+&\Theta(\smin-t_{Q1})\*\Theta(\smin-t_{12})\*\Theta(\smin-t_{2\Qb})
  \mbox{  1 and 2 soft}
\end{eqnarray}
The first line in the decomposition above corresponds to the resolved
region: both gluons are hard and not collinear to each other. 
Combined with the squared matrix element this contribution can be
integrated numerically in 4 dimensions.
The next three lines describe the single unresolved regions which yield
a singular contribution after phase space integration. There are also
single unresolved contributions when one of the two gluons becomes
`collinear' to the heavy quark or anti-quark. These contributions do not
yield a singular contribution after phase space integration, because of
the mass of the quarks. In principle they should be
of order $\smin$. In fact for small enough values for $\smin$ 
these contributions
vanish, because the quark mass serves already as a cutoff.
This can be checked by including them in the numerical
integration procedure. The last two lines describe `double unresolved'
contributions. As long as we restrict ourself to three-jet quantities,
these terms do not contribute.
The above slicing of the phase space differs slightly from the one used
in our previous work \cite{BeBrUw97a,BrUw97} and corresponds to the one used
in ref. \cite{KeLa99}. For the leading colour contributions the slicing 
(\ref{eq:theta}) is more convenient than the one used
in references \cite{BeBrUw97a,BrUw97}, because the soft factor (see below) 
is more compact and 
the separation of soft and collinear regions is simpler, facilitating 
the numerical implementation. We have checked numerically that the
two slicing methods are equivalent.

Let us now discuss the contributions from the soft regions, for
example the region where momentum $k_1$ is `soft':
\begin{eqnarray}
 \delta d\sigma_U= 
  \Theta(\smin-t_{Q1})\*\Theta(\smin-t_{12})\*\Theta(t_{2\Qb}-\smin)
  |S_1|^2 dR^d(k_Q,k_\Qb,k_1,k_2)
\end{eqnarray}
Using the soft factorization of the squared matrix element
\begin{eqnarray}
   N (N^2-1) |S_1|^2 \stackrel{\st{1 soft}}{\longrightarrow}
   g_s^2 \frac{N}{2} f(k_Q,k_1,k_2) |\T(Q(k_Q)\Qb(k_\Qb)g(k_2))|^2 
\end{eqnarray}
with 
\begin{eqnarray}
    f(k_i,k_s,k_j) 
    = 
    {4\,t_{ij}\over t_{is}t_{sj}}
    -{4\,k_i^2\over t_{is}^2}
    -{4\,k_j^2\over t_{js}^2},
\end{eqnarray}
together with the soft factorization of the phase space measure 
\begin{equation}
  dR^d(k_Q,k_\Qb,k_1,k_2) \to dR^d(k_Q,k_\Qb,k_2) 
  {1\over 2}\dRsoft
\end{equation}
(the factor 1/2 is the symmetry factor for the two identical gluons) 
we arrive at 
\begin{equation}
  \delta d\sigma_U = \frac{N}{2} S(m,0,t_{Q2}) 
  |\T(Q(k_Q)\Qb(k_\Qb)g(k_2))|^2 dR^d(k_Q,k_\Qb,k_2) 
\end{equation}
with 
\begin{equation}
  S(m_i,m_j,t_{ij}) = {g_s^2 \over 2} 
  \int \dRsoft f(k_i,k_s,k_j) \Theta(\smin-t_{is})\Theta(\smin-t_{sj}),
  \quad \mbox{for}\quad k_i^2 = m_i^2.
\end{equation}
Explicitly \cite{KeLa99},
\begin{eqnarray}
S(m,0,t_{Q2})&=& \frac{\alpha_s}{2\pi}\frac{1}{\Gamma(1-\e)}
\left(\frac{4\pi\mu^2}{\smin}\right)^{\e}
\left(\frac{t_{Q2}}{\smin}\right)^{\e}J(m,0,t_{Q2}),
\end{eqnarray}
where
\begin{eqnarray}
J(m,0,t_{Q2})
&=&\Theta(t_{Q2}-m^2)\left[
\frac{1}{\e^2}-\frac{1}{2\e^2}\left(\frac{t_{Q2}}{m^2}\right)^{\e}
+\frac{1}{2\e}\left(\frac{t_{Q2}}{m^2}\right)^{\e}
-\frac{1}{2}\zeta(2)+\frac{m^2}{t_{Q2}}\right] \nonumber \\
&+& \Theta(m^2-t_{Q2})\left(\frac{t_{Q2}}{m^2}\right)^{-\e}
\left[\frac{1}{2\e^2}+\frac{1}{2\e}-\frac{1}{2}\zeta(2)+1\right].
\end{eqnarray}
The case that gluon 2 becomes soft can be treated in the same way.
To finish our discussion of the singular contribution from $S_1$ let
us discuss the contribution from the collinear region.
Here the same procedure as in the massless case applies. For details
we refer to the previous section or to ref. \cite{GiGl92}. 
So we quote only the final result:
\begin{eqnarray}
  &&
  \Theta(t_{Q1}-\smin)\*\Theta(\smin-t_{12})\*\Theta(t_{2\Qb}-\smin)
  |S_1|^2 dR^d(k_Q,k_\Qb,k_1,k_2)\nn\\
  &\approx&
  N \cC_{gg}
  |\T(Q(k_Q)\Qb(k_\Qb)g(k_g))|^2 dR^d(k_Q,k_\Qb,k_g)
\end{eqnarray}
with 
\begin{eqnarray}
  \cC_{gg} 
  &=& 
  -{1\over 2} {g_s^2\over16\*\pi^2}\*{1\over\Gamma(1-\e)}
  \*\L{4\*\pi\mu^2\over \smin}\R ^{\e}
  {1\over \e}\int_{z_1}^{1-z_2} dz [z (1-z)]^{-\e} 
  P_{gg \to g}(z)\nn\\
  &=& 
  -{\as\over 2\*\pi}\*{1\over\Gamma(1-\e)}
  \*\L{4\*\pi\mu^2\over \smin}\R ^{\e}
  {1\over \e} I_{gg\to g},
\end{eqnarray}
and
\begin{eqnarray}
  I_{gg\to g} 
  &=& {(z_Q)^{-\e}\over \e } 
  + {(z_\Qb)^{-\e}\over \e }
  -{3\*(1-\e)\*(4-3\*\e)
    \over 2\* \e\*(3-2\*\e)}\*{\Gamma(1-\e)^2\over\Gamma(2-2\*\e)}\nn\\
 \end{eqnarray}
where $z_Q$ and $z_\Qb$ are defined as follows:
\begin{equation}
  z_Q = {\smin\over t_{Qg}},\qquad z_\Qb ={\smin\over t_{\Qb g}}.
\end{equation}
For $|S_2|^2$ we may apply the same procedure as for $|S_1|^2$. 
The soft singularities appearing in $|S_1+S_2|^2$ can be treated in a
similar way as the singularities in $S_1$. Since the amplitude $S_1+S_2$ 
is a QED-like  contribution with massive quarks, it 
does not induce collinear singularities. The soft singularities
can be isolated most easily by using the following identity: 
\begin{eqnarray}
  1=[\Theta(\smin-t_{Q1}-t_{\Qb1})+\Theta(t_{Q1}+t_{\Qb1}-\smin)]
  \*[\Theta(\smin-t_{Q2}-t_{\Qb2})+\Theta(t_{Q2}+t_{\Qb2}-\smin)].
\end{eqnarray}
If we expand this expression we obtain
\begin{eqnarray}
  1&=&-\Theta(\smin-t_{Q1}-t_{\Qb1})\*\Theta(\smin-t_{Q2}-t_{\Qb2})\nn\\
  &+& \Theta(\smin-t_{Q1}-t_{\Qb1})+\Theta(\smin-t_{Q2}-t_{\Qb2})\nn\\
  &+& \Theta(t_{Q1}+t_{\Qb1}-\smin)\*\Theta(t_{Q2}+t_{\Qb2}-\smin)
  \label{eq:subleading-slicing}
\end{eqnarray}
The first term describes a double unresolved contribution: both
gluons must be soft to satisfy the Heaviside functions. This term
does not contribute to a three-jet observable. The next two terms describe
the situation in which gluon 1 or gluon 2 is soft. It is still
allowed that both gluons are soft, this explains the minus sign in the
first line that avoids over-counting. 
The last term denotes the resolved contribution which 
can be calculated numerically in 4 dimensions.
Using the slicing given in \eq{eq:subleading-slicing} we thus obtain
for the soft contribution from $|S_1+S_2|$:
\begin{eqnarray}
 && [\Theta(\smin-t_{Q1}-t_{\Qb1})
  +\Theta(\smin-t_{Q2}-t_{\Qb2})]
  (-{1\over N}) (N^2-1)|S_1+S_2|^2 dR^d(k_Q,k_\Qb,k_1,k_2)\nn\\
  &&\to
  -{1\over N}S(m,m,t_{Q\Qb})  
  |\T(Q(k_Q)\Qb(k_\Qb)g(k_g))|^2 dR^d(k_Q,k_\Qb,k_g)
\end{eqnarray}
where we have used the factorization in the soft region which now takes
the form:
\begin{equation}
  -{1\over N}\*(N^2-1)\* |S_1+S_2|^2
  \stackrel{\st{1 soft}}{\longrightarrow}
  -{1\over 2 N} g_s^2 f(k_Q,k_1,k_\Qb) |\T(Q(k_Q)\Qb(k_\Qb) g(k_g))|^2.
\end{equation}
Note that we have once again identified the momentum of the remaining hard
gluon with $k_g$.
The explicit result for $S(m,m,t_{Q\Qb})$ is \cite{BrUw97}
\begin{eqnarray}
S(m,m,t_{Q\Qb})={\as \over 2\pi}{1\over \Gamma(1-\e)} 
  \left({4\pi\mu^2\over    \smin}\right)^{\e}
 \left({t_{Q\Qb}+2\*m^2\over    \smin}\right)^{\e}J(m,m,t_{Q\Qb}),
\end{eqnarray}
where
\begin{eqnarray}
J(m,m,t_{Q\Qb}) = {1\over \e} -{1+\omega\over 1-\omega}\ln(\omega)
+{1+\omega^2\over 1-\omega^2}\left\{{1\over \e}\*\ln(\omega)-
{1\over 2}\*\ln^2(\omega)-2\Li2(1-\omega)\right\} + O(\e)
\end{eqnarray}
and $\omega$ was defined in (\ref{eq:omega}).
Combining all singular contributions from $|S_1|^2$, $|S_2|^2$, and
$|S_1+S_2|^2$ we obtain the following result for the unresolved contribution:
\begin{eqnarray}
  &&
  \left\{ N S(m,0,t_{Qg}) + N S(m,0,t_{\Qb g}) + N \cC_{gg\to g}
  - {1\over N} S(m,m,t_{Q\Qb})
  \right\}
  d\sigma(Q(k_Q)\Qb(k_\Qb)g(k_g))\nn\\
  &=&  
  {\as \over 2\pi}{1\over \Gamma(1-\e)} 
  \left({4\pi\mu^2\over \smin}\right)^{\e} 
  \times  
  \Bigg\{ 
  N \left[ \L {t_{Qg}\over \smin}\R ^\e J(m,0,t_{Qg}) 
  + \L {t_{\Qb g}\over \smin}\R ^\e J(m,0,t_{\Qb g}) - {1\over \e} I_{gg\to g}
  \right]
  \nn\\
  &-& {1\over N}\left({t_{Q\Qb}+2\*m^2\over    \smin}\right)^{\e}
 J(m,m,t_{Q\Qb})
  \Bigg\} d\sigma(Q(k_Q)\Qb(k_\Qb)g(k_g))
  \label{eq:qqgg-singular}
\end{eqnarray}
To get the singular 
(unresolved) contribution to $h_6$ one has simply to replace 
$d\sigma(Q(k_Q)\Qb(k_\Qb)g(k_g))$ by $\h6born $. Note that both 
quantities $d\sigma(Q(k_Q)\Qb(k_\Qb)g(k_g))$ and $\h6born $ are
evaluated in $d$ dimensions. As we will show in the next subsection
the same structure is obtained from the virtual corrections. 
After the cancellation of the IR singularities 
we thus need $d\sigma(Q(k_Q)\Qb(k_\Qb)g(k_g))$ or
equivalently $\h6born$ only in 4 dimensions.

\subsection{Infrared and collinear finiteness}
\label{sec:IR-cancellation}
We close this section by demonstrating that the singular contributions
discussed in the previous subsections indeed cancel the singularities 
encountered in the virtual corrections.

Combining \eq{eq:IR-div-LC}, \eq{eq:IR-div-SC}, and \eq{eq:ren-contr}
we obtain for the IR singularities of the virtual corrections the 
following result:
\begin{eqnarray}
  \label{eq:v-singular}
  h_6^{\st{virt., IR div.}}&=&  
  {\as\over 2\*\pi} \* \rg
  \L {4\*\pi\*\mu^2  \over m^2 } \R ^{\e}
  \Bigg\{
  -N 
  \LB
  {1\over \e^2}
  - {1\over \e}\* \ln({t_{Qg}\over m^2})
  - {1\over \e}\* \ln({t_{\Qb g}\over m^2}) \RB
  \nn\\
  &&
  +{1\over N}{1\over \e}\*
  {1+\omega^2\over 1-\omega^2}\*\ln(\omega)-{1\over \e}\LB2  C_F 
  - {1\over 6} 
  \*(2\* n_f^l -11\* N)\RB
  \Bigg\} \*\h6born
\end{eqnarray}

The singular parts of the real emission processes are given by
(\eq{eq:QQqq-singular}, \eq{eq:qqgg-singular}):
\begin{eqnarray}
  \label{eq:r-singular}
  \left.h_6^{\st{real, unres.}}\right|_{\st{div}}
  &=&{\as \over 2\pi}{1\over \Gamma(1-\e)} 
  \left({4\pi\mu^2\over m^2}\right)^{\e} 
 \nn\\
  &\times & 
  \Bigg\{ 
  N \left[ {1\over \e^2}
  -{1\over \e}\ln\L {t_{Qg}\over m^2}\R
  -{1\over \e} \ln\L {t_{\Qb g}\over m^2}\R\right]
  \nn\\ &&
  -{1\over N}  {1\over \e}
    {1+\omega^2\over 1-\omega^2}\*\ln(\omega)
    +{1\over\e }\*\left[2 C_F - {1\over 6}\* (2 n_f^l -11 N )\right]
  \Bigg\}\*\h6born,
\end{eqnarray}
where we have used the expansions
\begin{eqnarray}
  \label{eq:expansions}
  \L {t_{Qg}\over \smin}\R ^\e J(m,0,t_{Qg}) 
  &=& 
  {1\over 2\e^2}
  + {1\over 2\e} 
  -{1\over 2\e}\ln\L {\smin\over m^2}\R + j(m,0,t_{Qg})+O(\e) \nn\\
  \left({t_{Q\Qb}+2\*m^2\over    \smin}\right)^{\e}  J(m,m,t_{Q\Qb})
  &=&
  {1\over \e} + {1\over \e}\*{1+\omega^2\over 1-\omega^2}\*\ln(\omega)
  + j(m,m,t_{Q\Qb })+O(\e),\nn\\
  I_{gg\to g}
  &=&
  -{11 \over 6}-\ln({\smin\over t_{Qg}})-\ln({\smin\over t_{\Qb g}})\nn\\
  &+&({1 \over 2} \*\ln^2({\smin\over t_{Qg}})
  +{1 \over 2} \*\ln^2({\smin\over t_{\Qb g}})
  +{1 \over 3} \*\pi^2
  -{67 \over 18} )\*\e + O(\e^2),\nn\\
  n_f^l {1\over \e}I_{q\qb\to g}&=& 
  n_f^l \*{1\over \e} \* {1 \over 3} +{5 \over 9}n_f^l +O(\e),
\end{eqnarray}
with $j(m,0,t_{Qg}), j(m,m,t_{Q\Qb })$ denoting the order 1 contributions
in the expansion. They are given by
\begin{eqnarray}\label{eq:jlc}
j(m,0,t_{Qg}) &=& {1\over 4}\*\ln^2\L{\smin\over m^2}\R
-{1\over 2}\*\ln\L{\smin\over m^2}\R+1-{1\over2}\zeta(2) \nn \\
&+&\Theta(t_{Qg}-m^2)\*\left\{
-{1\over 2}\ln^2\L{t_{Qg}\over m^2}\R+\ln\L{t_{Qg}\over m^2}\R
-1+{m^2\over t_{Qg}} \right\},
\end{eqnarray}
\begin{eqnarray}\label{eq:jsc}
j(m,m,t_{Q\Qb })& =&  \ln\L{t_{Q\Qb}+2\*m^2\over    \smin}\R
\left\{1+ {1+\omega^2\over 1-\omega^2}\ln(\omega)\right\}
-{1+\omega\over 1-\omega}\ln(\omega) \nn \\
&-&{1+\omega^2\over 1-\omega^2}\left\{
{1\over 2}\*\ln^2(\omega)+2\*\Li2(1-\omega)\right\}.
\end{eqnarray}
Note that we have not expanded the factor $\mu^\e$ to show explicitly
the cancellation of this factor as it was promised in section 
\ref{sec:renormalization}. 
Comparing \eq{eq:v-singular} and \eq{eq:r-singular} one sees immediately 
that the singular contribution from the real corrections cancel exactly the
IR divergences from the virtual corrections.

\section{Summary and conlusions}
\label{sec:conclusions}
In this work we have calculated the virtual corrections to 
the parity-violating functions $F_3$ and $F_6$ of the fully
differential cross section \eq{eq:dsigma}
for $e^+e^-\to Q\bar{Q}g$. 
We have given results for the UV and IR singularities as well as
for the finite contributions. We have shown explicitly the cancellation
of the UV singularities after the renormalization has been performed. 
In addition we have calculated also the singular contribution from real 
corrections using the phase space slicing method. We checked that this
contribution cancels exactly the IR singularities in the virtual
corrections. Together with our comparison to the known results for
massless
quarks the cancellation of the IR and UV singularities serves as an
important check of our calculation.

We have given our next-to-leading order results 
in terms of one function $h_6$, from which $F_3$ and $F_6$ 
can be reconstructed
using \eq{eq:f36}--\eq{eq:h7}.
The final result for the contribution from three resolved
partons, $h_6^{\st{3 res., NLO}}$, can be obtained from
the following formula:
\begin{equation}
 h_6^{\st{3 res., NLO}}  = 
  h_6^{\st{virt., fin.}}+h_6^{\st{rest}}
  + \left.h_6^{\st{real, unres.}}\right|_{\st{fin.}},
\end{equation}
where $ h_6^{\st{virt., fin.}}$ is given in \eq{eq:h6NLOfin}.
The contribution $ h_6^{\st{rest}}$ is given  by
\begin{eqnarray}
  h_6^{\st{rest}}&=& h_6^{\st{virt., UV div.}}
  +
  \left.h_6^{\st{virt., IR div.}}\right|_{\st{lead. colour}} 
  +
  \left.h_6^{\st{virt., IR div.}}\right|_{\st{subl. colour}} \nn\\
  &+&
  \left.h_6^{\st{virt., ren.}}\right|_{\st{UV}}
  +
  \left.h_6^{\st{virt., ren.}}\right|_{{\gamma_5}}
  + \left.h_6^{\st{real, unres.}}\right|_{\st{div.}}\nonumber \\
 &=&
  8\*\as^2\* (N^2-1)\*C_F\* B\*\bigg( 
  1+x-2\*\xb+(1-\xb)\*{\xb\over x}\nn\\
  &+&z\*\bigg[2\*B\*
  ((\xb^2-2\*x)\*(1-x)+(2-3\*x)\*(1-\xb))
  +g_1\*{x^2-4\*z\over x\*(1-x)}
  +4\*z\*B\*(2\*{(1-\xb)^2\over 1-x}+\xg)\bigg]
  \bigg)\nn\\
  &+& 16\*\as^2\*C_F \*B\*(
  {1\over \xg}\*(x\*\xb-4\*\xb+2+\xb^2)
  -2\*z\*\xg\*B\*(x-\xb)
  )\nn\\ 
  &+& 
  {\as\over 4\*\pi}\*  
  \left\{ {1\over 3}\*(2\* n_f -11\* N)\ln({m^2\over \mu^2})
  +{2\over 3}\*\sum_i \ln{m_i^2\over m^2} - 8  C_F\right\}  \h6born
  - 32\as^2   (N^2-1) C_F
  \deltah6
  -\frac{\as}{\pi} C_F \h6born,\nn\\
\end{eqnarray}
where $g_1$ was defined in (\ref{eq:g_1def}).
The finite contribution from unresolved partons is given by
\begin{eqnarray}
  \label{eq:j'en-ai-marre}
  \left.h_6^{\st{real, unres.}}\right|_{\st{fin.}}
  &=&
  {\as \over 2\pi}
  \Bigg\{ 
  N \bigg[ j(m,0,t_{Qg}) 
  +  j(m,0,t_{\Qb g}) 
  - ({1 \over 2} \*\ln^2({\smin\over t_{Qg}})
  +{1 \over 2} \*\ln^2({\smin\over t_{\Qb g}})
  +{1 \over 3} \*\pi^2
  -{67 \over 18} )
  \bigg]
  \nn\\
  &-& {1\over N} 
   j(m,m,t_{Q\Qb})-{5\over 9}n_f^l
  \Bigg\} \h6born.
\end{eqnarray}
The functions $j(m,0,t_{Qg})$, and $j(m,m,t_{Q\Qb})$ are  
given in \eq{eq:jlc} and \eq{eq:jsc}, respectively. Note that
$\left.h_6^{\st{real, unres.}}\right|_{\st{fin.}}$ depend on the method
used to combine virtual and real corrections. Furthermore we can set
$d=4$ in $\h6born$ in \eq{eq:j'en-ai-marre} because all 
singularities are cancelled.

The results presented in this paper comprise the whole contribution
of three `resolved partons'. Parity-violating three-jet observables
can be computed using these results together with the 
contribution from four `resolved partons'. The latter 
does not contain IR singularities and can, 
for a given jet algorithm, be calculated numerically in a
rather straightforward
fashion using known results of the matrix elements for the 4-parton final
states \cite{BaMaMo94}.
In particular one can determine  the 3-jet and 4-jet  forward-backward
asymmetries
$A_{\st{FB}}$ for massive quark jets to order $\alpha_s^2$ in the 
QCD coupling, for instance for $b$ quark jets at the $Z$ peak.

\bibliographystyle{revtex}
\bibliography{literatur}
\end{document}